# Blue shifting of the A exciton peak in folded monolayer 1H-MoS$_2$

[1,*,†]Frank J. Crowne, [1,†]Matin Amani, [1,†]A. Glen Birdwell, [1]Matthew L. Chin, [1]Terrance P. O'Regan, [2]Sina Najmaei, [2]Zheng Liu, [2]Pulickel M. Ajayan, [2]Jun Lou, and [1]Madan Dubey

[1]Sensors and Electron Devices Directorate, US Army Research Laboratory, Adelphi MD 20723, USA

[3]Department of Mechanical Engineering and Materials Science, Rice University, Houston, TX 77005, USA

The large family of layered transition-metal dichalcogenides is widely believed to constitute a second family of two-dimensional (2D) semiconducting materials that can be used to create novel devices that complement those based on graphene. In many cases these materials have shown a transition from an indirect bandgap in the bulk to a direct bandgap in monolayer systems. In this work we experimentally show that folding a 1H molybdenum disulphide (MoS$_2$) layer results in a turbostratic stack with enhanced photoluminescence quantum yield and a significant shift to the blue by 90 meV. This is in contrast to the expected 2H-MoS$_2$ band structure characteristics, which include an indirect gap and quenched photoluminescence. We present a theoretical explanation to the origin of this behavior in terms of exciton screening.

**Keywords:** Molybdenum disulfide, 2D semiconductors, photoluminescence, folds

* Author to whom correspondence should be addressed. e-mail: frank.j.crowne2.civ@mail.mil
† Equally contributing authors.

Transition-metal dichalcogenide compounds, in particular molybdenum disulfide (MoS$_2$), have become the focus of considerable interest in the past few years due to their great potential as a complementary material to graphene in advanced high performance flexible/transparent digital electronics and optoelectronic circuits as well as for use in new quantum mechanical devices [1, 2]. Bulk samples of MoS$_2$ are indirect-gap semiconductors, with an indirect bandgap of 1.3 eV. This arises from the presence of two conduction band minima (CBM) at different points in the Brillouin zone, one of which is at the K-point directly above the valence band maximum (VBM) and another of lower energy between the K and Γ points. The bandgap is thus the energy difference between the lower energy CBM and the VBM, making it indirect.

Bulk MoS$_2$ exists in two polytypes: hexagonally stacked 2H-MoS$_2$ and tetragonal 1T-MoS$_2$. All bulk samples discussed in this manuscript are initially of the 2H polytype, which is made up of 6- layer units in the stacking order S-Mo-S-S-Mo-S, with the second three layers rotated by 180° relative to the first and weakly coupled by van der Waals (vdW) interactions between the two S-Mo-S layers [3]. As the sample thickness is reduced down to a few atomic layers, the indirect bandgap is observed to widen. Band-structure calculations suggest this widening is due to a rapid increase in the energy of the lower CBM while the higher energy CBM and VBM move only slightly. For layer thicknesses down to a 2H monolayer, the lower CBM determines the bandgap, and the bandgap remains indirect. Because the S-S bonds in monolayer MoS$_2$ are weakly coupled, exfoliation allows this monolayer to be further divided, resulting in a three atomic layer sheet with a simple S-Mo-S stacking [4, 5]. In this paper we adopt the nomenclature of Ataca et al. [3] and identify these layers as the 1H-MoS$_2$ polytype, which unlike the 2H polytype, no longer has a center of inversion. Remarkably, in 1H-MoS$_2$ the two CBM trade places in terms of energy, leaving the K-point CBM directly above the VBM, which is now increased by approximately 100 meV, and converting the bandgap for a single 1H layer from indirect to direct. 1H-MoS$_2$ is the first 2D material artificially created by exfoliating a 3D indirect-gap material to exhibit an induced indirect to direct bandgap transition. This result has lead to numerous studies that focused on using this material to fabricate optoelectronic devices [6, 7].

Like graphene, layers of 1H-MoS$_2$ have very little resistance to bending and can be subjected to strains greater than ~10% without failure [8]. Also like graphene, these layers can be folded onto themselves to form coupled two-layer structures [9, 10]. Among the possible arrangements of folded two layers include some that mimic a 2H-MoS$_2$ monolayer (fig. 1). In this paper we discuss the photoluminescence (PL) and Raman spectra of folded 1H-MoS2 layers and characterize their electrical properties. We report that, in contrast to a true 2H stacked monolayer, a turbostratic (T-MoS$_2$) bilayer, can exhibit PL emission which is blue shifted by approximately 90 meV, along with an enhanced photoluminescence quantum yield (QY) compared to an unfolded 1H layer [11]. As progress is made towards forming complex heterostructures using 2D materials, it is important to understand the effects of rotational misalignment, poor interlayer coupling, or general turbostratic behavior on the band structure. This is especially true for dichalcogenides systems such as MoS$_2$, because the change from 2H to 1H results in a fundamental change in the materials electronic properties.

Bulk MoS$_2$ films were grown directly by CVD onto a 285 nm SiO$_2$/Si substrate using the procedure described in detail by Ajayan et. al [14, 21]. In brief, high aspect ratio MoO$_3$ nanoribbons were grown using hydrothermal processes, and then dispersed onto an auxiliary silicon substrate and placed inside a furnace tube with the growth substrate and sulfur powder. The furnace was heated to a peak temperature of 850°C under a constant flow of nitrogen. Folded regions were created on these MoS$_2$ crystals by spinning a PMMA layer on the substrate and removing it in flowing acetone. This resulted in the stripping of a small fraction of the crystallites from the substrate and a smaller number being folded over onto the crystal. E-beam lithography (EBL) was then used to fabricate back-gated FETs on these folded flakes of 1H-MoS$_2$. The MoS$_2$ layers were patterned using a low power ICP etch in a CH$_4$/O$_2$ plasma and e-beam evaporated Ti/Au (15/85 nm) was used to form source and drain contacts.

High-resolution Raman and photoluminescence (PL) imaging were performed after each processing step. These measurements were performed with a WITec Alpha 300RA system using the 532 nm line of a frequency-doubled Nd:YAG laser as the excitation source. The spectra were measured in the backscattering configuration using a 100x objective and either a 600 or 1800 grooves/mm grating. The spot size of the laser was ~342 nm resulting in an incident laser power density of ~140 μW/μm$^2$. Raman imaging found no time dependent shifting of the E$_{2g}$ and A$_{1g}$ modes of 2H-MoS$_2$ layers or the E' and A$_1$' modes of 1H-MoS$_2$ layers during testing [3]. In addition, atomic force microscopy (AFM) and scanning electron microscopy (SEM) images were taken to verify layer count/height, and to search for residues left during processing. The field-effect mobility of folded-layer and monolayer regions was characterized using a Keithley 4200 semiconductor characterization system by using the substrate as a universal back-gate with all measurements taken under high vacuum.

Optical and scanning electron microscope (SEM) images as well as atomic force microscope (AFM) scans are shown in figure 2 for a typical region around a fold. The step heights measured via AFM for the monolayer regions (fig. 2e) show a slightly greater thickness (0.9 nm) than what is normally observed for 1H layer material prepared by exfoliation (0.7 nm), whereas the measured 2H step height (fig. 2g) is in very good agreement with the literature value. Despite the disagreement between the thickness measurements for exfoliated and CVD 1H-$MoS_2$ material there is a strong consensus in the literature that this material is in fact high-quality monolayer based upon independent evidence derived from Raman, PL, and X-TEM [14]. We believe that the thickness discrepancy we see is caused either by imperfect bonding between the $SiO_2$ and $MoS_2$ or carbon contamination at the $MoS_2$-$SiO_2$ interface; this can be observed in cross sectional TEM images shown in previous reports on growth of this material [15]. The step height at the edge of the folded region is 1.85 nm, which again may be due to imperfect bonding to the substrate as shown in figures 2d and 2e. The RMS roughness measured for both the monolayer and folded-layer regions is 2.7 Å and 2.4 Å respectively (measured $SiO_2$ roughness was 1.3 Å), which suggests that we have a defect free $MoS_2$ to $MoS_2$ interface. We provide several complete sets of metrology results, showing additional examples of folded regions obtained both on this sample and on other samples in the supplementary materials section; these highlight the reproducibility of the blue shift and increased PL intensity for the folded $MoS_2$ material.

High resolution Raman and PL intensity/position maps of the folded region are shown in figures 3 and 4 respectively. In addition, the average of x = 60 spectra obtained with long integration times for the 1H layer, folded layer, and 2H monolayer regions are shown in figure 5. For the monolayer spectra we can observe an average peak separation of 21.6 cm$^{-1}$ and a A'/E' intensity ratio of 1.8, which is in good agreement reports on high quality monolayer CVD grown MoS$_2$ reported previously in the literature. For the case of the folded 1H-MoS$_2$ layer we observe a hardening of the in-plane E' peak in the neighborhood of the fold, indicating the presence of compressive strain [8]. The same trends are observed along the grain boundary defects in the MoS$_2$ crystal located directly above the fold. The corresponding PL mapping shows a highly uniform 90 meV blue shift of the A exciton peak in the folded region with no significant reduction in the QY compared to unfolded-layer material. It is unlikely that the blue shift in the folded region is a result of compressive strain or doping due to the uniformity of the shift compared to the complex compressive/tensile strain distribution that can be anticipated for a fold [12]. It is important to note that the uniformity of the PL intensity is very dependent on the sample growth conditions, which can lead to uniform, edge enhanced, defect enhanced, or quenched behavior [13, 14, 15]. Recent literature has been published on edge enhanced PL in CVD grown monolayer WS$_2$, claiming that the increased QY in the PL signal is due to the accumulation and radiative-recombination of bound excitons at defects along the sample edge leading to a red shift in the PL spectra [13]. However, the authors acknowledge that the behavior and the proposed mechanisms need more experimental and theoretical analysis.

We also fabricated a back-gated field effect transistor on the folded MoS$_2$ region and compared the results with a device having the same geometry, but fabricated on a 1H layer within the same sample. Images of the device at various stages of the fabrication process are shown in figure 6 as well as I$_{DS}$-V$_{BG}$ transfer curves for the two devices (both with a length (L) = 800 nm and width (W) = 1μm) plotted on both logarithmic and linear scales. We see an expected increase in the current for the folded sample which is likely due to the increase in the sample thickness, leading to a higher conductivity in saturation. Both the device built on the fold as well as the device built on a nearby pristine flake show very similar field effect mobilities of 1.12±0.15 $cm^2/V·s$ and 0.95±0.2 $cm^2/V·s$ respectively, which is reasonable for devices built on $SiO_2$ with no top dielectric (typical back gated field effect mobility values range from 8 to 0.5 $cm^2/V·s$) [16]. In the folded samples, the pinch-off voltage was significantly more negative, indicating that the folded region has a higher donor dopant density [17].

The presence of a blue shift in the PL of this system is remarkably hard to explain, in that many potential causes, e.g., strain or quantum confinement, turn out to produce red shifts instead [17]. After much discussion, we focused our attention on excitons as possible enablers of the blue shift. Under illumination, free carriers generated in the course of PL measurements are usually assumed to equilibrate with a sizable population of excitons. In $MoS_2$ layers these excitons are transitional between Frenkel-type and Wannier-type, and highly localized since both the conduction and valence bands are built primarily out of orbitals from the Mo atoms [18, 19]. Their motion is thus diffusive throughout the structure and has no effect on, e.g., device electrical characteristics. In addition, they are weakly screened in the direction perpendicular to the plane, so that the Coulomb force between electron and hole is stronger in this direction, while the in-plane screening weakens these forces. As a result, excitons are flattened perpendicular to the Mo plane and spread out parallel to it. A folded structure created from a single 1H-$MoS_2$ layer brings a pair of monolayer sheets into close proximity. Figure 1 shows this process schematically. We note that in general the sheets will be misoriented; however, for folds along certain high-symmetry directions can result in a 6-layer configuration with precisely the structure of a 2H-MoS$_2$, including the inversion of one sheet relative to the other. Because the band gap of this monolayer is indirect and smaller than that of the monolayer, we can infer that for this special orientation of the fold a red shift should appear in the PL when the sheets of the fold are very close. The question that then arises is: could a more remote separation or a turbostratic orientation of the sheets lead to a blue shift? This line of inquiry led us to the model we describe below.

Figure 5 shows a comparison of the experimental Raman and PL curves for a 1H-MoS$_2$ layer (blue), a folded 1H-MoS$_2$ layer (red), and a 2H-MoS$_2$ monolayer (green). Since the peaks shown are actually generated by exciton recombination light, we base our theoretical explanation on the recombination kinetics of these excitons. As a result of overall PL pumping, each sheet of the fold has its own population of excitons, which we label $Ex_1$ and $Ex_2$ for the top and bottom sheet, respectively. As the sheets approach one another, the excitons interact across the physical gap via Van der Waals (VdW) forces. When the distance between the layers is large enough, the space between the sheets is free of charge with all the charge localized near the sheets. This implies that the fields in between the sheets are governed by Laplace's equation using crystal-field theory, and hence can be expanded in multipole series whose structure is determined by the point-group symmetries of the material. This expansion can be derived by exploiting translation symmetry, which calls for a 2D Fourier expansion of the charge density $\rho(\vec{X}, z)$ of the 2D layer:

$$\rho(\vec{X},z) = \frac{(2\pi)^2}{\Omega} \sum_{\vec{j}} \int_{\Omega} \rho(\vec{\Xi},z) e^{-i\vec{\Gamma}_{\vec{j}} \cdot (\vec{X}-\vec{\Xi})} d\vec{\Xi}. \tag{1}$$

Here $\vec{\Gamma}_{\vec{j}}$ are 2D reciprocal lattice vectors (RLVs) and $\Omega$ is the area of a unit cell of sheet material; note that the integral over $\vec{\Xi}$ is confined to a single unit cell. To obtain the corresponding (3D) electrostatic potential we use the Poisson equation:

$$\nabla^2 \Phi(\vec{X},z) = \left(\frac{\partial^2}{\partial \vec{X}^2} + \frac{\partial^2}{\partial z^2}\right) \Phi(\vec{X},z) = \frac{1}{\varepsilon} \rho(\vec{X},z) \tag{2}$$

Although the density is zero outside the sheets, the potential is nonzero. Expanding this potential in the same way as the charge density:

$$\Phi(\vec{X},z) = \sum_{\vec{j}} e^{-i\vec{\Gamma}_{\vec{j}} \cdot \vec{X}} F_{\vec{j}}(z) \tag{3}$$

We can then find the z-dependent Fourier coefficients $F_{\vec{j}}(z)$ of the potential:

$$F_{\vec{j}}(z) = \frac{1}{2\Gamma_{\vec{j}}} \frac{1}{\varepsilon} \frac{1}{\Omega} \int_{\Omega} e^{i\vec{\Gamma}_{\vec{j}} \cdot \vec{\Xi}} d\vec{\Xi} \int_{-\infty}^{\infty} d\zeta e^{-|\vec{\Gamma}_{\vec{j}}||z-\zeta|} \rho(\vec{\Xi},\zeta) \tag{4}$$

If the region between the sheets is wide enough, only the longest-range fields are needed to describe the field from one sheet at the position of the other, i.e., those fields associated with the smallest $\vec{\Gamma}_{\vec{j}}$:

$$\Phi(z) \approx C_0 e^{-|\vec{\Gamma}_0| z} \tag{5}$$

where $\vec{\Gamma}_0$ is the smallest reciprocal lattice vector and

$$C_0 = \frac{1}{2|\vec{\Gamma}_0|} \frac{1}{\varepsilon} \frac{1}{\Omega} \int_\Omega d\vec{\Xi} e^{i\vec{\Gamma}_0 \cdot \vec{\Xi}} \int_{-\infty}^{\infty} d\zeta e^{-|\vec{\Gamma}_0| \zeta} \rho(\vec{\Xi}, \zeta) \tag{6}$$

is the field strength. For high-symmetry arrangements of the atoms within the sheets the integrals over $\vec{\Xi}$ will vanish and with them the fields of the smaller RLV; thus, a high-symmetry graphene sheet will have only short-range VdW fields arising from the higher-order RLV. This is especially significant for $MoS_2$, in that the structure of a 2H-$MoS_2$ (six-layer) monolayer possesses a center of inversion, in contrast to the 1H-$MoS_2$ (three-layer) structure which does not; thus, long-range VdW fields will be present outside a 1H-$MoS_2$ sheet that are absent from a 2H-$MoS_2$ sheet. If the sheets are misoriented with respect to one another this will be enhanced due to the total loss of symmetry.

The existence of anomalously long-range electrostatic forces between turbostratic 1H-$MoS_2$ sheets in close proximity has important implications for the kinetic theory of excitonic populations in a folded structure. Because excitons couple coherently to photons to form polaritons, recombination by photon emission is strongly inhibited [19]; however, this is not true for recombination by exciton-exciton collisions, which are by nature incoherent. Conservation of energy implies reactions of the form

$$Ex + Ex \rightarrow Ex^* + \Phi, \tag{7}$$

where $Ex^*$ is an exciton in an excited state and $\Phi$ denotes either photons or phonons. These reactions will occur as long as $2\mathcal{E}_{ex} > \mathcal{E}_{ex}^* + \mathcal{E}_\Phi$, where $\mathcal{E}_{ex}, \mathcal{E}_{ex}^*$ are exciton binding energies and $\mathcal{E}_\Phi$ is the energy of the residual photons or phonons. More problematic is the reaction:

$$Ex + Ex \rightarrow e^- + h^+ + \Phi \qquad (8)$$

where $e^-$ is a free electron, $h^+$ is a free hole, and $\Phi$ is a long-wavelength photon or multiple phonons, since energy conservation requires that $2\mathcal{E}_{ex} > \mathcal{E}_G$. For Wannier excitons this inequality is easily satisfied, but for MoS$_2$ it is less obvious, since some theoretical estimates predict values of $\mathcal{E}_{ex}$ approaching 1 eV [22] versus a value of 1.9 eV for $\mathcal{E}_G$.

Nevertheless, if we postulate that (8) can take place, and the two excitons are in the same sheet, the force responsible for this collisional process should be dipolar in nature, and therefore inhibited by free-carrier screening. Suppose, however, that the excitons are in different sheets, so the reaction is

$$Ex_1 + Ex_2 \rightarrow \begin{cases} e_1^- + h_2^+ + \Phi \\ e_2^- + h_1^+ + \Phi \end{cases} \qquad (9)$$

Then the anisotropic shape of excitons in 1H-MoS$_2$ suggests that the fields they carry may extend into the gap like the VdW fields described above, making them available to mediate the process (9). Because the screening in a 2D layer is anisotropic and weak in the direction perpendicular to the plane, these fields are anisotropic as well; specifically, they are enhanced in the inter sheet region relative to in-plane fields. Since the gap is too wide for electrons and holes to tunnel across, the reaction products add to the screening free charge in both sheets. This in turn reduces the binding energy of the remaining excitons, leading to a blue shift in the recombination energy. If this blue shift is larger than the red shift of the free carrier absorption edge, we will observe an overall blue shift in the PL. Figure 7 shows a possible diagram for process (9). Note that this diagram requires the non-tunneling transfer of a hole from one sheet to another in order to complete the annihilation of the lower exciton $Ex_2$.

In summary we report on the experimental observations of folded CVD grown 1H-MoS$_2$ monolayers. In marked contrast to the transition from an indirect to direct bandgap material that occurs when a 2H-MoS$_2$ monolayer is converted into a 1H-MoS$_2$ layer, we observe that the reverse process of folding a 1H-MoS$_2$ layer to form a nominally 2H-MoS$_2$ monolayer does not lead to conversion back to an indirect-gap material and a red shift of the PL accompanied by a loss of QY. Rather, folding results in an anomalous enhanced photoluminescence coupled with a blue shift in the A exciton peak by 90 meV. We have proposed a theoretical explanation for our experimental observation, which is able to explain the uniformity of the PL blue shift over the area of the fold and is insensitive to whether the sheets are in register or are rotated relative to one another. This is analogous to turbostratic stacking in graphene bilayers, and provides some insight into behaviors we can anticipate from turbostratic MoS$_2$ films. However, the 2D exciton behavior in two dimensional dichalcogenides must be further investigated in order to develop advanced heterostructures utilizing 2D interlayer interactions with these materials.

## Acknowledgements

The authors acknowledge the support of the U.S. Army Research Laboratory (ARL) Director's Strategic Initiative (DSI) program on interfaces in stacked 2D atomic layered materials. The authors would also like to thank Dr. Pani Varanasi, ARO for his in-depth technical discussion on 2D atomic layers R&D.  P.M.A., J.L., S.N, and Z.L. also acknowledge funding support from the ARO MURI program on 2D materials. The views and conclusions contained in this document are those of the authors and should not be interpreted as representing the official policies, either expressed or implied, of the ARL or the U.S. Government. The U.S. Government is authorized to reproduce or distribute reprints for Government purposes notwithstanding any copyright notation herein.

## List of Figures

**Fig. 1:** Schematic of a single 1H-MoS$_2$ layer (a), single 2H-MoS$_2$ monolayer (b), and a folded 1H-MoS$_2$ layer (c). Note the structural similarity of the latter to a 2H-MoS$_2$ monolayer far from the connection region.

**Fig. 2:** Optical (a), secondary electron (b), backscatter (c), and AFM (d) images of folded MoS$_2$, as well as corresponding scans from 1H layer (e), folded 1H layer (f), and 2H monolayer regions (g).

**Fig. 3:** High resolution maps of Raman peak intensity (a, c) and position (b, d) of the in plane (a, b) and out of plane (c, d) modes in a folded 1H-MoS$_2$ layer.

**Fig. 4:** High resolution maps showing the PL intensity (a) and peak position (b) in folded 1H-MoS$_2$.

**Fig 5:** Raman (a) and PL (b) spectrum for a 1H-MoS$_2$ monolayer (blue), folded 1H-MoS$_2$ (red), and a 2H-MoS$_2$ monolayer (green).

**Fig 6:** Optical micrographs depicting the fabrication flow of a back-gated FET on folded 1H-MoS$_2$ (a, b, c): application of photoresist over the folded region (a), PMMA patterned to deposit ohmic contacts on the etched region (b), and the finished device (c). $I_{DS}$-$V_{BG}$ characteristics for both 1H-MoS$_2$ (red) and folded 1H-MoS$_2$ (blue) are plotted in both log and linear scales for a L = 800 nm, W = 1 µm FET measured using the 2 point probe technique (d).

**Fig 8:** Feynman diagram for intersheet exciton-exciton collision. Exciton *eh* pairs are denoted by squares, the diamond shows a *eh*-photon/phonon vertex, and the triangle shows intersheet *h-h* elastic scattering due to fold edge. Red lines delineate the time ordering of this particular process. The symbol × denotes the nondispersive (energy conserving) conversion of hole $h_1$ into hole $h'_2$, i.e., transfer of a hole from one sheet to another, most likely by non-tunneling injection through the fold connection region.

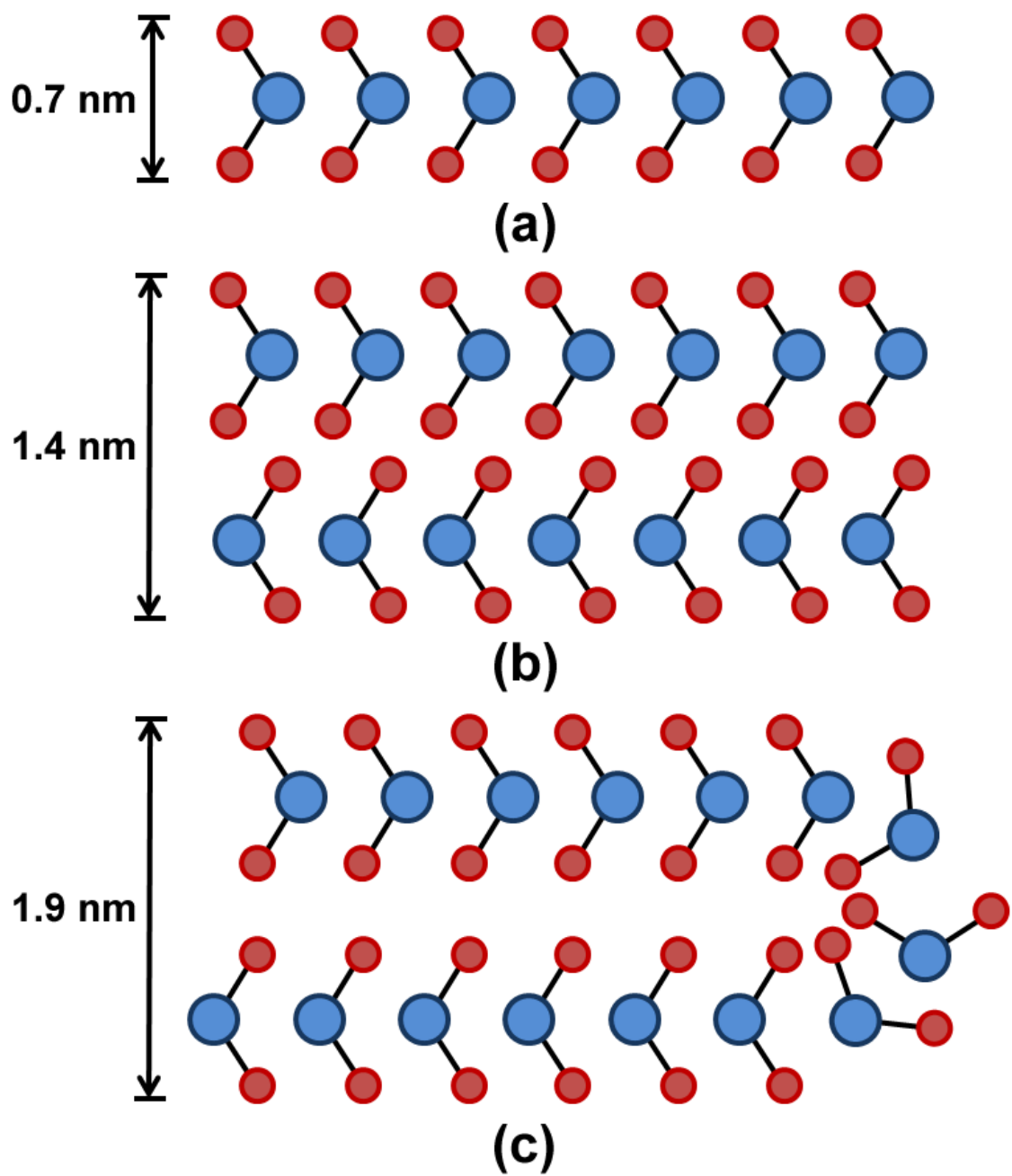

**Fig. 1:** Schematic of a single 1H-MoS$_2$ layer (a), single 2H-MoS$_2$ monolayer (b), and a folded 1H-MoS$_2$ layer (c). Note the structural similarity of the latter to a 2H-MoS$_2$ monolayer far from the connection region.

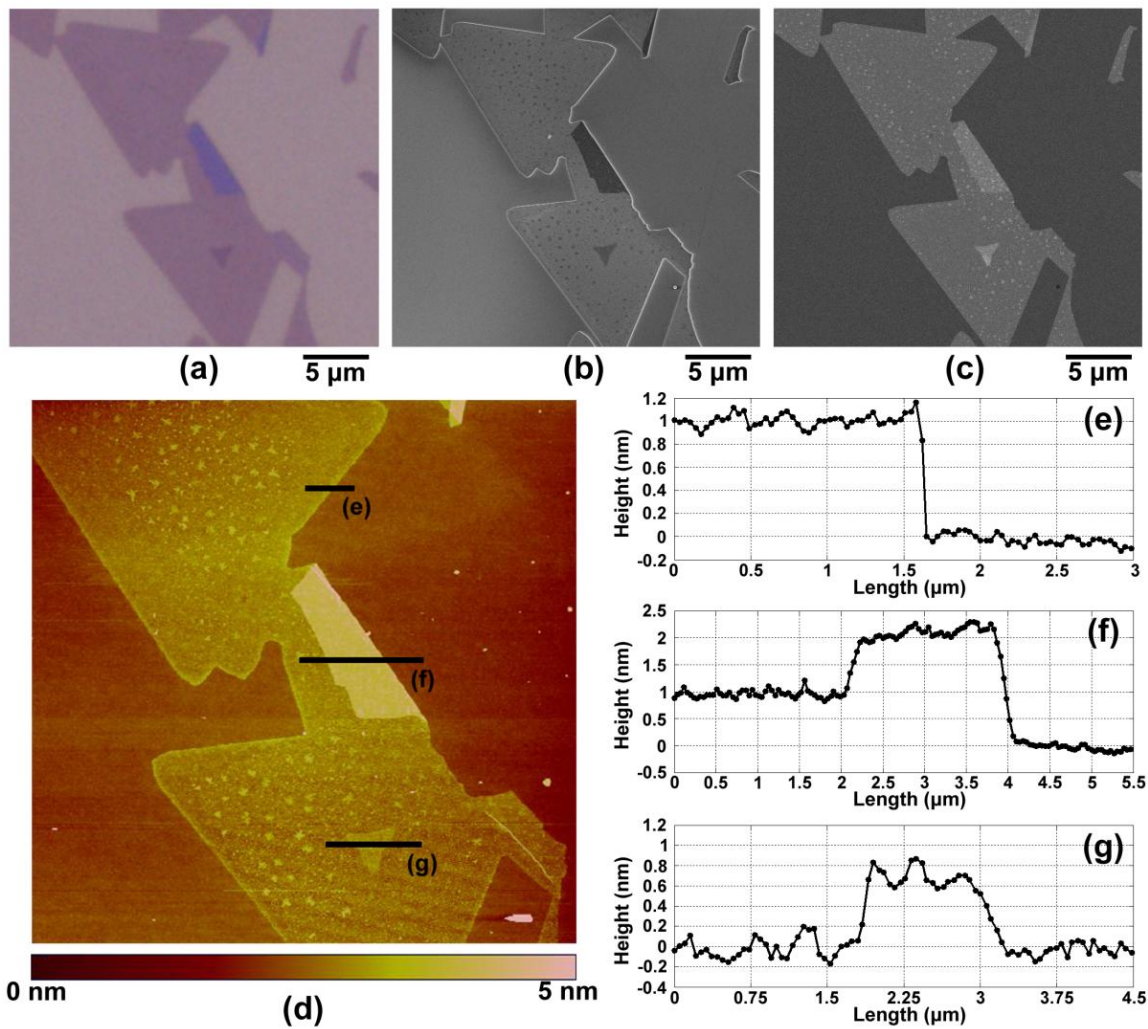

**Fig. 2:** Optical (a), secondary electron (b), backscatter (c), and AFM (d) images of folded MoS$_2$, as well as corresponding scans from 1H layer (e), folded 1H layer (f), and 2H monolayer regions (g).

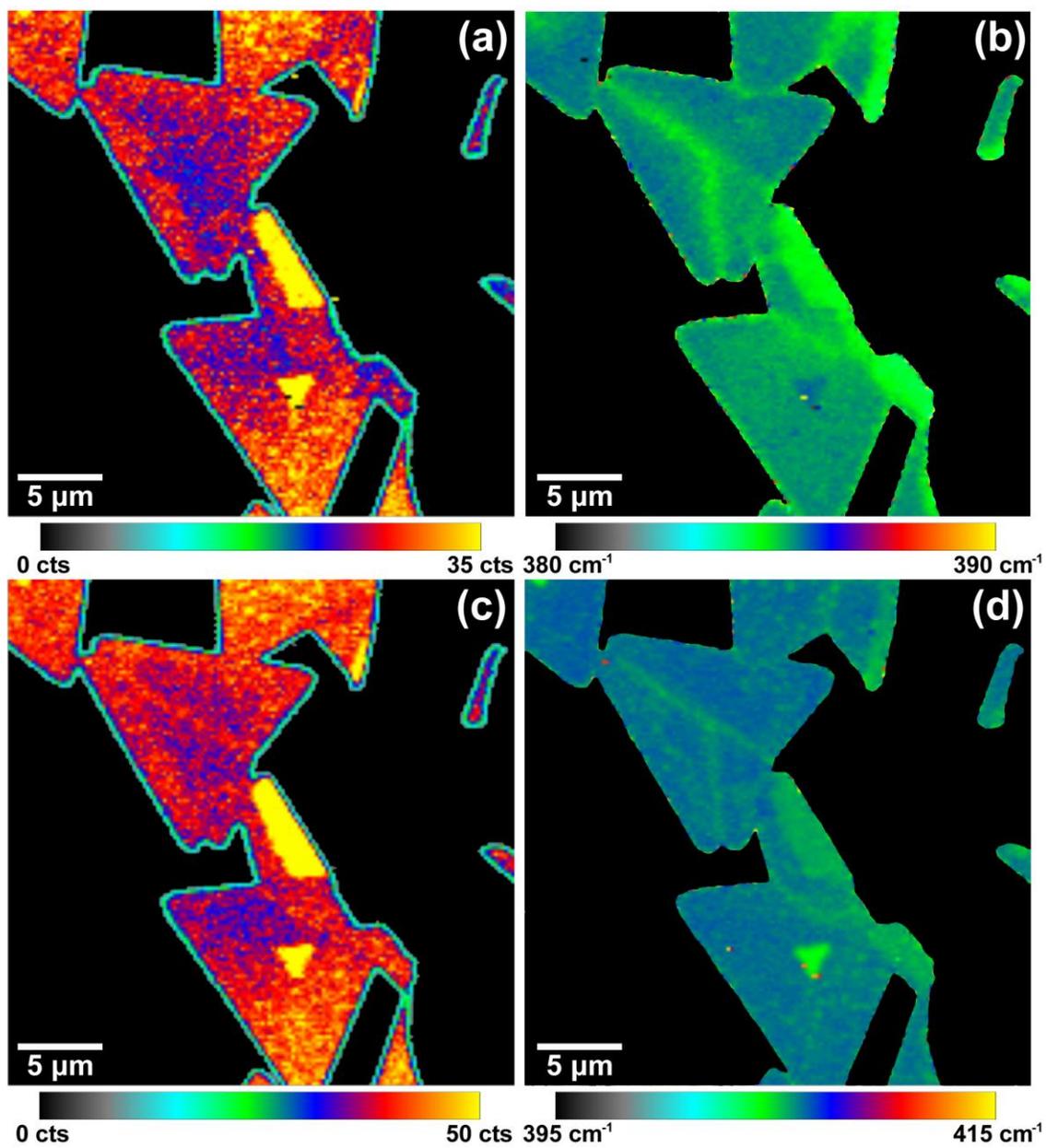

**Fig. 3:** High resolution maps of Raman peak intensity (a, c) and position (b, d) of the in plane (a, b) and out of plane (c, d) modes in a folded 1H-MoS$_2$ layer.

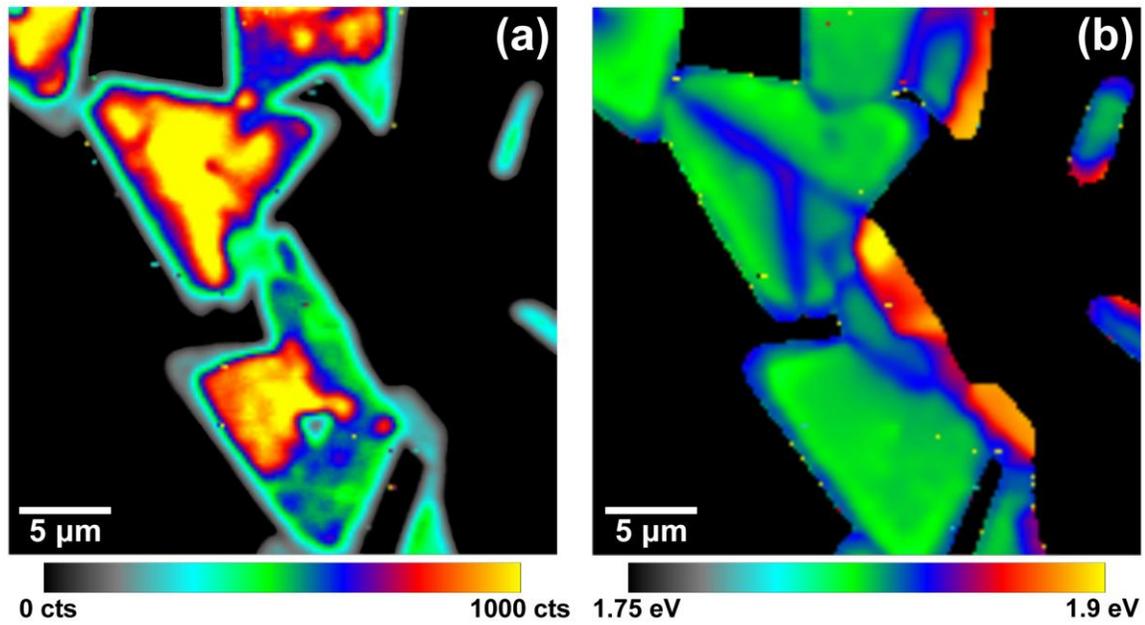

**Fig. 4:** High resolution maps showing the PL intensity (a) and peak position (b) in folded 1H-$MoS_2$.

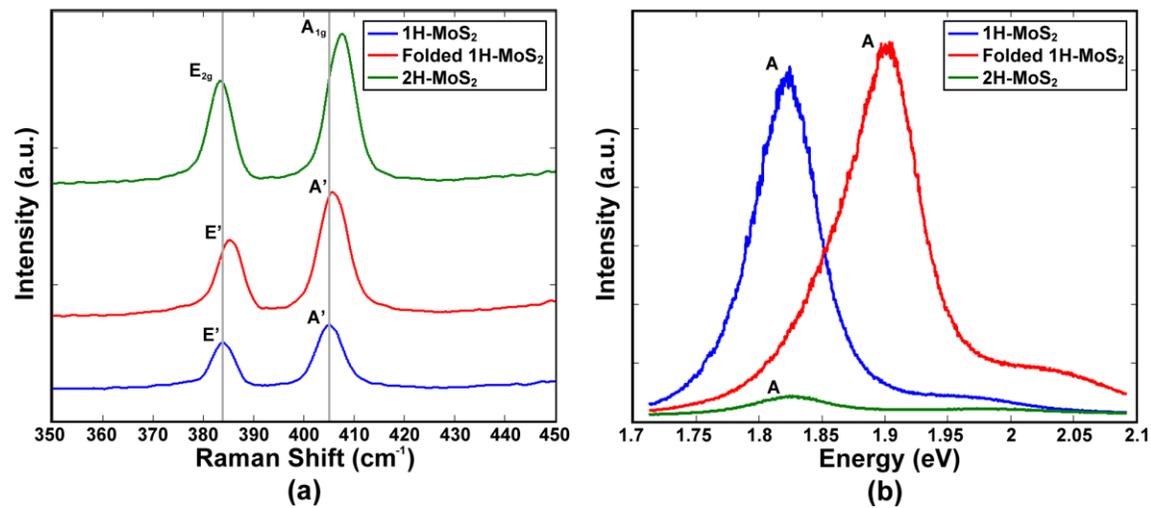

**Fig 5:** Raman (a) and PL (b) spectrum for a 1H-MoS$_2$ monolayer (blue), folded 1H-MoS$_2$ (red), and a 2H-MoS$_2$ monolayer (green).

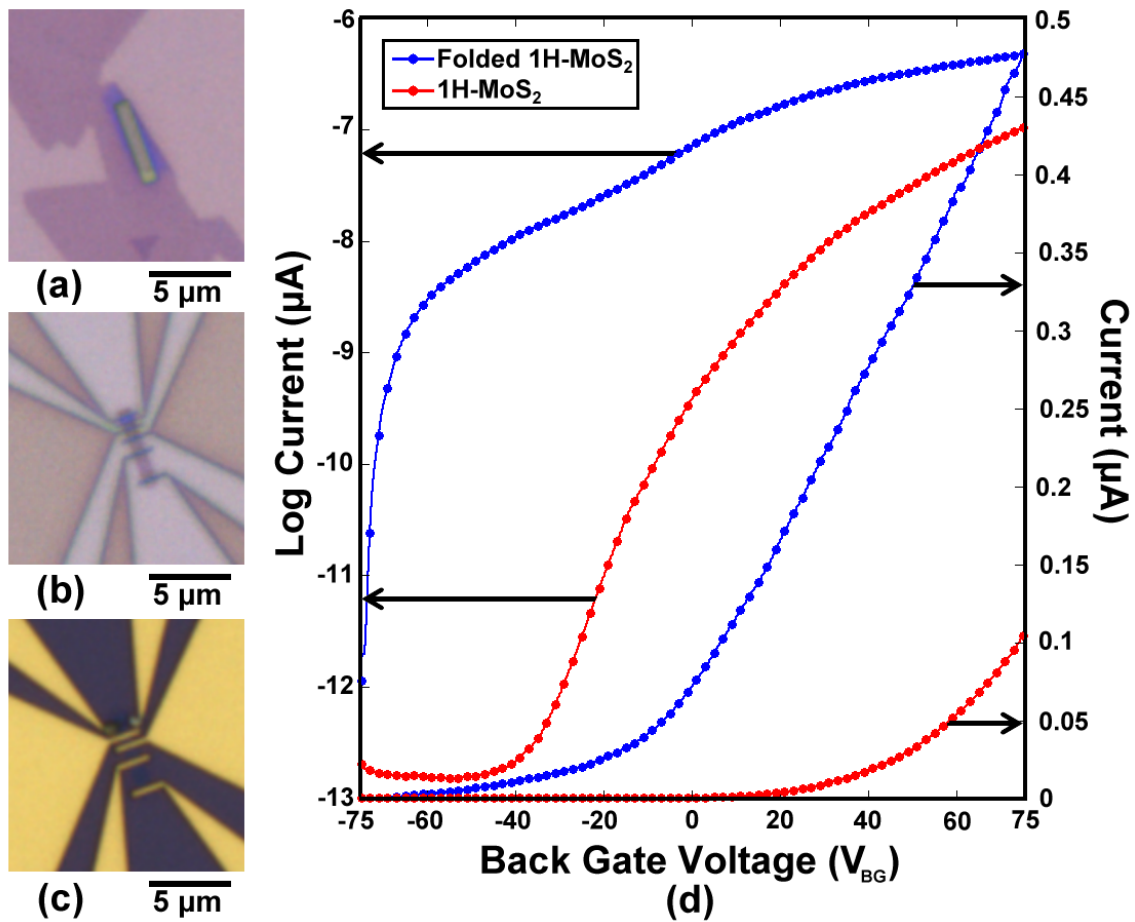

**Fig 6:** Optical micrographs depicting the fabrication flow of a back-gated FET on folded 1H-MoS$_2$ (a, b, c): application of photoresist over the folded region (a), PMMA patterned to deposit ohmic contacts on the etched region (b), and the finished device (c). I$_{DS}$-V$_{BG}$ characteristics for both 1H-MoS$_2$ (red) and folded 1H-MoS$_2$ (blue) are plotted in both log and linear scales for a L = 800 nm, W = 1 μm FET measured using the 2 point probe technique (d).

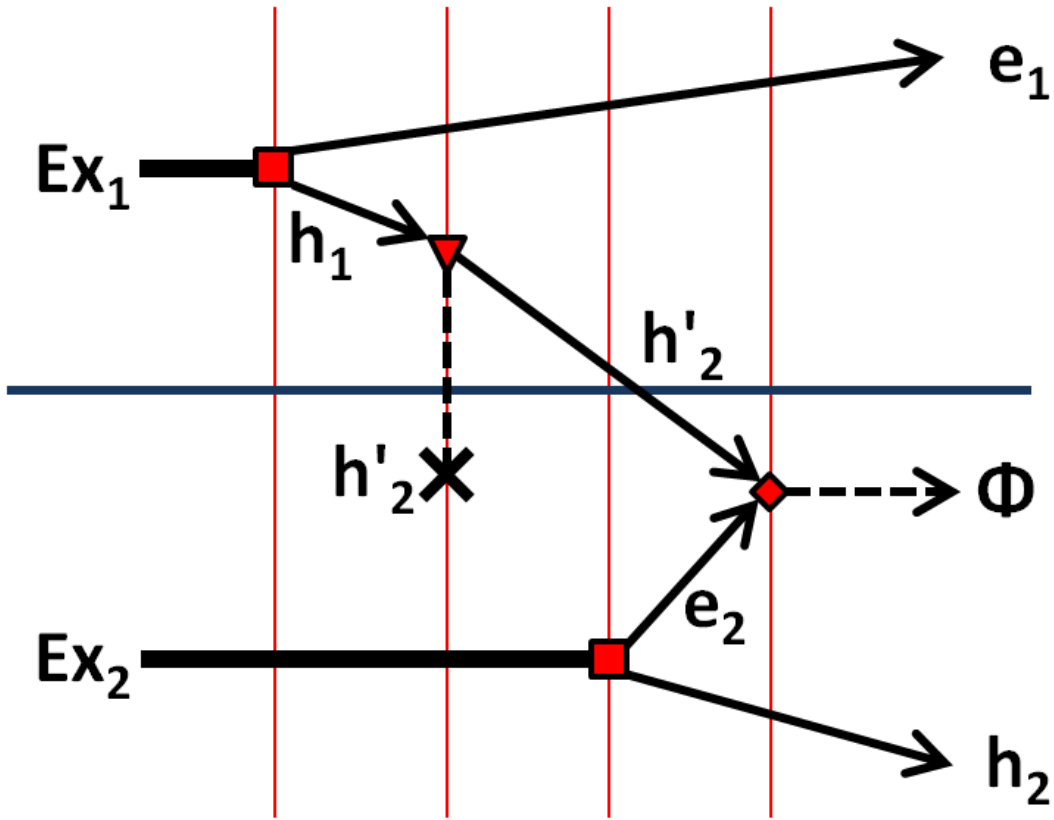

**Fig 8:** Feynman diagram for intersheet exciton-exciton collision. Exciton *eh* pairs are denoted by squares, the diamond shows a *eh*-photon/phonon vertex, and the triangle shows intersheet *h-h* elastic scattering due to fold edge. Red lines delineate the time ordering of this particular process. The symbol × denotes the nondispersive (energy conserving) conversion of hole $h_1$ into hole $h'_2$, i.e., transfer of a hole from one sheet to another, most likely by non-tunneling injection through the fold connection region.